# Self-Propagating High-Temperature Synthesis of Boron Phosphide

Vladimir A. Mukhanov,[1] Petr S. Sokolov,[1] Yann Le Godec,[2] and Vladimir L. Solozhenko [1,*]

[1] *LSPM–CNRS, Université Paris Nord, 93430 Villetaneuse, France*

[2] *IMPMC, Université P&M Curie, 75005 Paris, France*

**Abstract** – A new method of producing boron phosphide (BP) submicron powders by self-propagating high-temperature reaction between boron phosphate and magnesium in the presence of an inert diluent (sodium chloride) has been proposed. Bulk polycrystalline BP with microhardness of $H_V = 28(2)$ GPa has been prepared by sintering the above powders at 7.7 GPa and 2600 K.

Boron phosphide, BP, is a wide band gap $A^{III}B^{V}$ semiconductor compound with a diamond-like structure [1] that is characterized by high thermal and chemical stability (up to 1500 K in air [2]) and high hardness ($H_V \approx 30$ GPa [3,4]). The main limitation for the wide use of BP is the lack of relatively simple and economical methods of its production.

Boron phosphide can be synthesized either by direct interaction of the elements [1,2] or by reaction between halogenides of boron and phosphorus in the presence of sodium [5,6]; while BP single crystals may be grown by the crystallization from flux solutions [7,8] or by gas-transport reactions in two-zone furnaces [1,9]. The disadvantages of these methods are: the use of toxic and aggressive reagents; rather complicated technical implementation, high labour intensity and time consumption. Here we propose a new simple and rapid method of producing BP submicron powders using readily available and cheap reagents.

The basis of our method is the reduction of boron phosphate with magnesium i.e.

$$BPO_4 + 4\,Mg = BP + 4\,MgO \qquad (1)$$

that has not been described before. The possibility of this reaction to occur as self-propagating high-temperature synthesis (SHS) has been studied. Amorphous boron phosphate ($BPO_4$) produced by the procedure described in [10] and magnesium metal (> 99.5%, 315/200 μm) were used as reagents. The $BPO_4$ and Mg were mixed in 1:4.1 molar ratio (a small excess of magnesium) and pressed in a steel die at a load corresponding to the pressure of 0.6 GPa into pellets 40 mm in diameter and 20 mm high (experimental density 1.5–1.6 g/cm³). To conduct the reaction, a pellet was placed onto a substrate from pressed MgO, the center of its upper surface was heated to ~1000 K by a sharp flame of a gas burner thus initiating the SHS process, and covered with an alumina crucible to prevent the oxidation of magnesium by atmospheric oxygen. In the course of reaction (1) a large quantity of heat is released, and the front of the reaction proceeds at a mean velocity ~ 2 mm/s in the vertical and 1 mm/s in radial directions; the time of complete combustion of a pellet is 20-30 s. After cooling the resulting loosely held compacts were crushed, treated with distilled water, and the residue was boiled for an hour in excess of 5N-hydrochloric acid, and then many times washed with distilled water, and dried in air at 50°C.

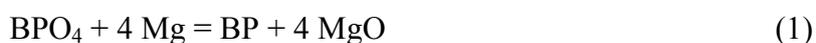

X-ray diffraction (XRD) analysis of the products was performed on an Equinox 1000 Inel powder diffractometer (Co$K\alpha$1 radiation, $\lambda$ = 1.789007 Å), and their morphology was studied on a Supra 40 VP Carl Zeiss high resolution scanning electron microscope. Raman spectra were excited by 632.8 nm He-Ne laser (beam size ~10 μm) and recorded on a Horiba Jobin Yvon HR800 micro-Raman spectrometer.

The washed products of reaction (1) were boron phosphide with impurity (up to 30%) of boron subphosphide, $B_{12}P_2$. The presence of $B_{12}P_2$ is caused by a high (> 1500 K) temperature of the reaction mixture combustion that results in a partial decomposition of the as-forming BP [1,2]. To reduce the temperature in the reaction front, we used sodium chloride, a chemically inert diluent, ($T_{melt}$ = 1074 K). An addition of NaCl to the initial reaction mixture was accompanied by a decrease in the intensity of the $B_{12}P_2$ diffraction lines in the reaction products, and as the NaCl content attains 50 wt %, the formation of the almost single-phase (> 98%) boron phosphide (Fig. 1a) with the lattice parameter of $a$ = 4.5356(9) Å, which is close to the literature value (4.537 Å [11]), and a mean particle size of 100–200 nm (Fig. 2a) was observed. The Raman spectra of the washed reaction (1) products (Fig. 3a) exhibit two features: strong asymmetric line at ~828 cm$^{-1}$ and weak broad line at ~800 cm$^{-1}$ that are characteristic bands for BP [12]. In some points of the samples we observed weak sharp line at ~476 cm$^{-1}$ (the most intense band of $B_{12}P_2$ [13]).

The yield of boron phosphide BP makes about 35% of theoretical yield by reaction (1), which stems from a number of side reactions (formation of magnesium phosphide, boron and phosphorus oxides, etc.). However, the relatively low yield of the desired product is compensated by the simplicity of the method and availability of the reagents. The higher (> 50 wt %) NaCl content results in incomplete combustion of a reaction mixture, which leads to a decrease of the BP yield.

Washed powders of single-phase boron phosphide were treated at 7.7 GPa and 2600 K for 3 min in a high-temperature cell of a toroid-type high-pressure apparatus. The experimental details are described elsewhere [14]. The recovered samples were single-phase polycrystalline bulks of boron phosphide with the lattice parameter of $a$ = 4.5349(9) Å (Fig. 1b) and grain size of 5–10 μm (Fig. 2b). Only characteristic lines of BP are present in the Raman spectra after high pressure – high temperature treatment (Fig. 3b), in good agreement with X-ray diffraction data that additionally proves the phase purity and homogeneity of the samples.

The Vickers hardness was measured using a Duramin-20 (Struers) microhardness tester at 10 N load and indentation time of 10 s. According to the data obtained, polycrystalline bulk boron phosphide has microhardness of 28(2) GPa which approaching that of BP single crystals [3].

The authors thank Dr. Ovidiu Brinza for electron-microscopy studies and Dr. Thierry Chauveau for assistance in XRD analysis. This work was financially supported by the Agence Nationale de la Recherche (grant ANR-2011-BS08-018).

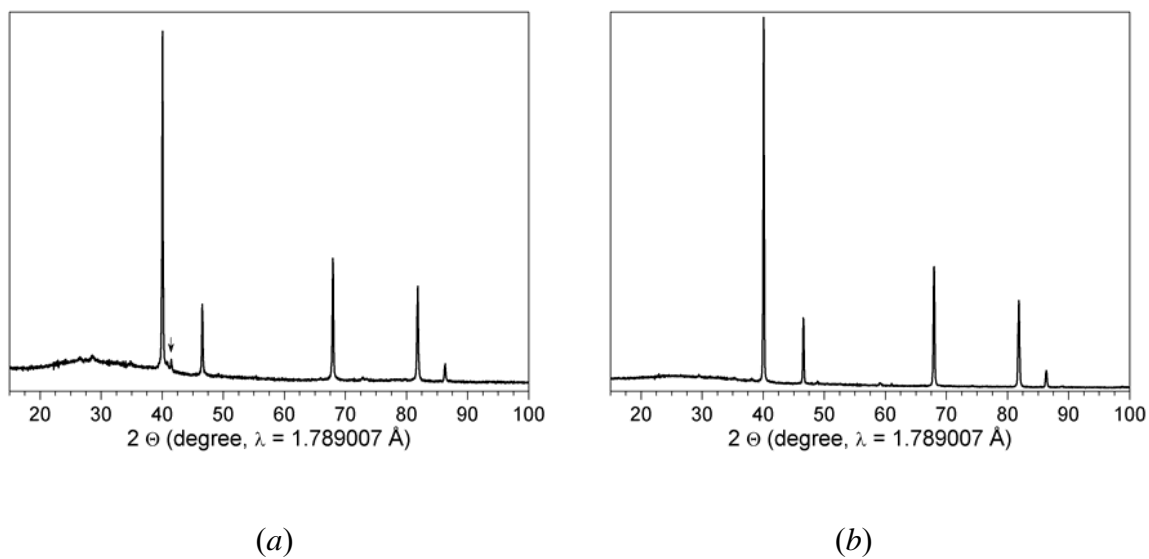

**Fig. 1.** Diffraction patters of the washed sample of boron phosphide produced by SHS (*a*) and of the same sample after treatment at 7.7 GPa and 2600 K (*b*). The arrow shows the location of 104 and 021 diffraction lines of $B_{12}P_2$.

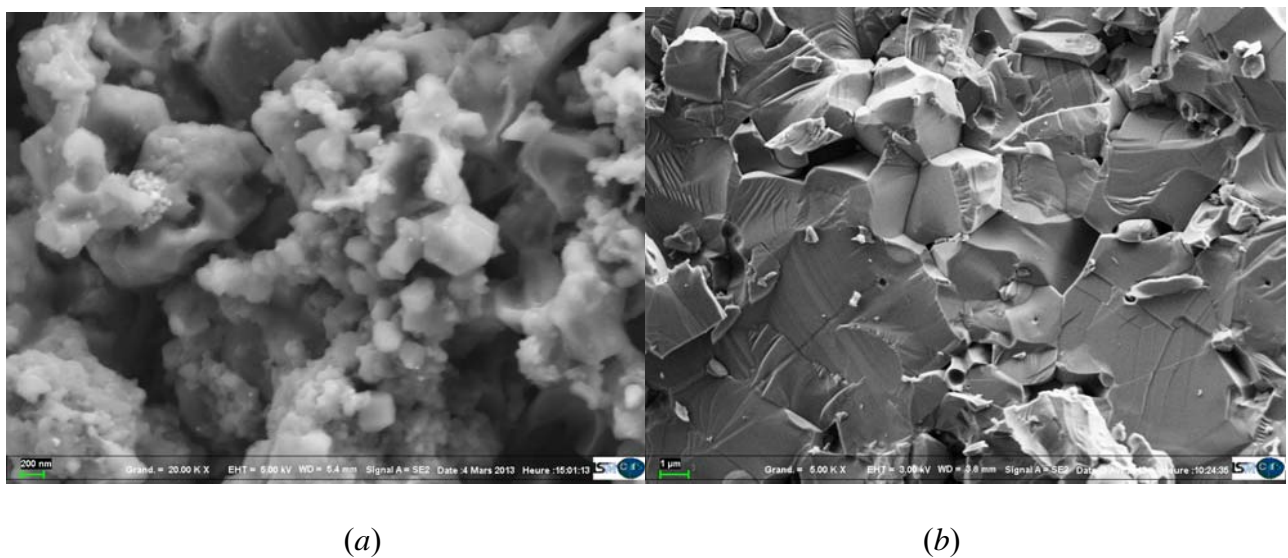

**Fig. 2.** SEM micrographs of the washed sample of boron phosphide produced by SHS (*a*) and of the same sample after treatment at 7.7 GPa and 2600 K (*b*).

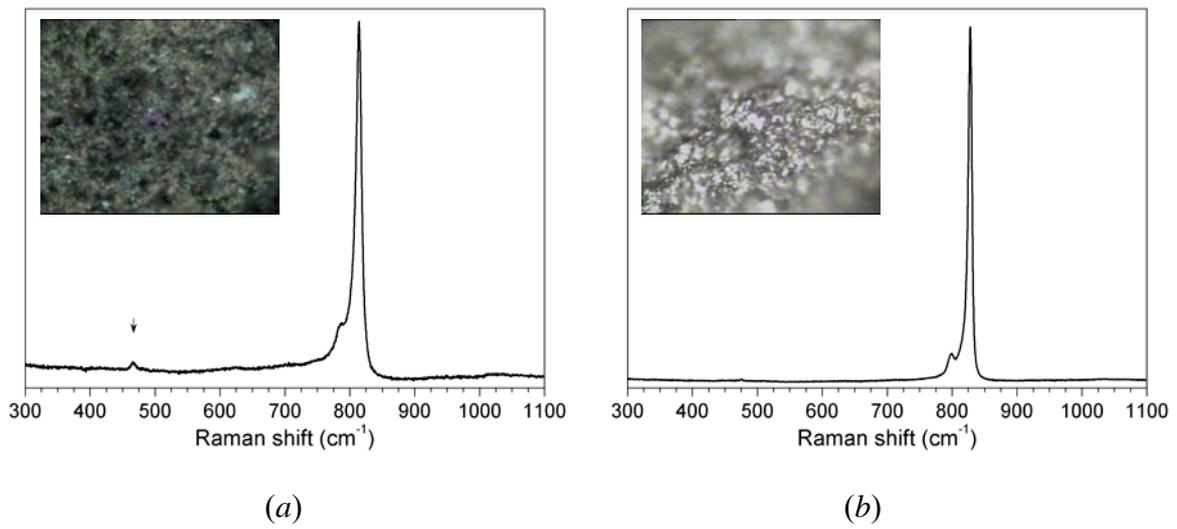

**Fig. 3** Raman spectra of the washed sample of boron phosphide produced by SHS before (*a*) and after treatment at 7.7 GPa and 2600 K (*b*). The arrow points to the characteristic band of $B_{12}P_2$ (478 cm$^{-1}$). Insets show optical images (100×) of the sample surfaces.